\documentclass{sig-alternate}
\usepackage{url}
\usepackage{algpseudocode, hyperref}
\usepackage{subfigure}
\begin{document}
\makeatletter
\def\@copyrightspace{\relax}
\makeatother

%

\title{Uncovering the Perfect Place: Optimising Workflow Engine Deployment in the Cloud}
%
%
%
%
%

\numberofauthors{1} 
%
\author{
%
%
\alignauthor
Michael Luckeneder and Adam Barker\\
       \affaddr{School of Computer Science}\\
       \affaddr{University of St Andrews, UK}\\
       \email{adam.barker@st-andrews.ac.uk}
}

\date{30 July 1999}

\maketitle
\begin{abstract}

When orchestrating highly distributed and data-intensive Web service workflows the geographical placement of the orchestration engine can greatly affect the overall performance of a workflow. We present CloudForecast: a Web service framework and analysis tool which, given a workflow specification, computes the optimal Amazon EC2 Cloud region to automatically deploy the orchestration engine and execute the workflow.  We use geographical distance of the workflow, network latency and HTTP round-trip time between Amazon Cloud regions and the workflow nodes to find a ranking of Cloud regions. This overall ranking predicts where the workflow orchestration engine should be deployed in order to reduce overall execution time. Our experimental results show that our proposed optimisation strategy, depending on the particular workflow, can speed up execution time on average by 82.25\% compared to local execution.

\end{abstract}





\section{Introduction}
Scientific workflows are typically orchestrated using a workflow engine running locally within an organisation's network. However, if the Web services in the workflow are data-intensive and spread across many geographical regions, the data might have to move long distances in order to flow from the data sources to the Web services via the orchestrator. This in turn most likely slows down the execution of the workflow and degrades the overall performance. 

A possible solution could be to use an Infrastructure as a Service (IaaS) Cloud, such as an Amazon EC2 instance, it is possible to automatically deploy the orchestrator into a suitable EC2 region that is ``closer'' to the data source and Web service nodes; in turn data would not have to travel as far and therefore execution times could be reduced.

We design, implement and evaluate CloudForecast, a pre-deployment analysis tool\footnote{\url{https://github.com/bigdatalab/movingdata}}  which can dynamically compute the ``optimal'' Cloud region to deploy the orchestrator, given a specific workflow consisting of multiple distributed services. We focus on Directed Acyclic Graph (DAG) based workflows since these are heavily used in the scientific community \cite{DBLP:journals/tsc/BarkerWH12, DBLP:journals/jsw/BesanaPBRG09}. In order to characterise the ``optimal'' Cloud region, we use total geographical distance, network latency and HTTP round-trip time of the workflow.



\

\section{Architecture}
\begin{figure}[!htp]
\centering
\includegraphics[width=2.3in]{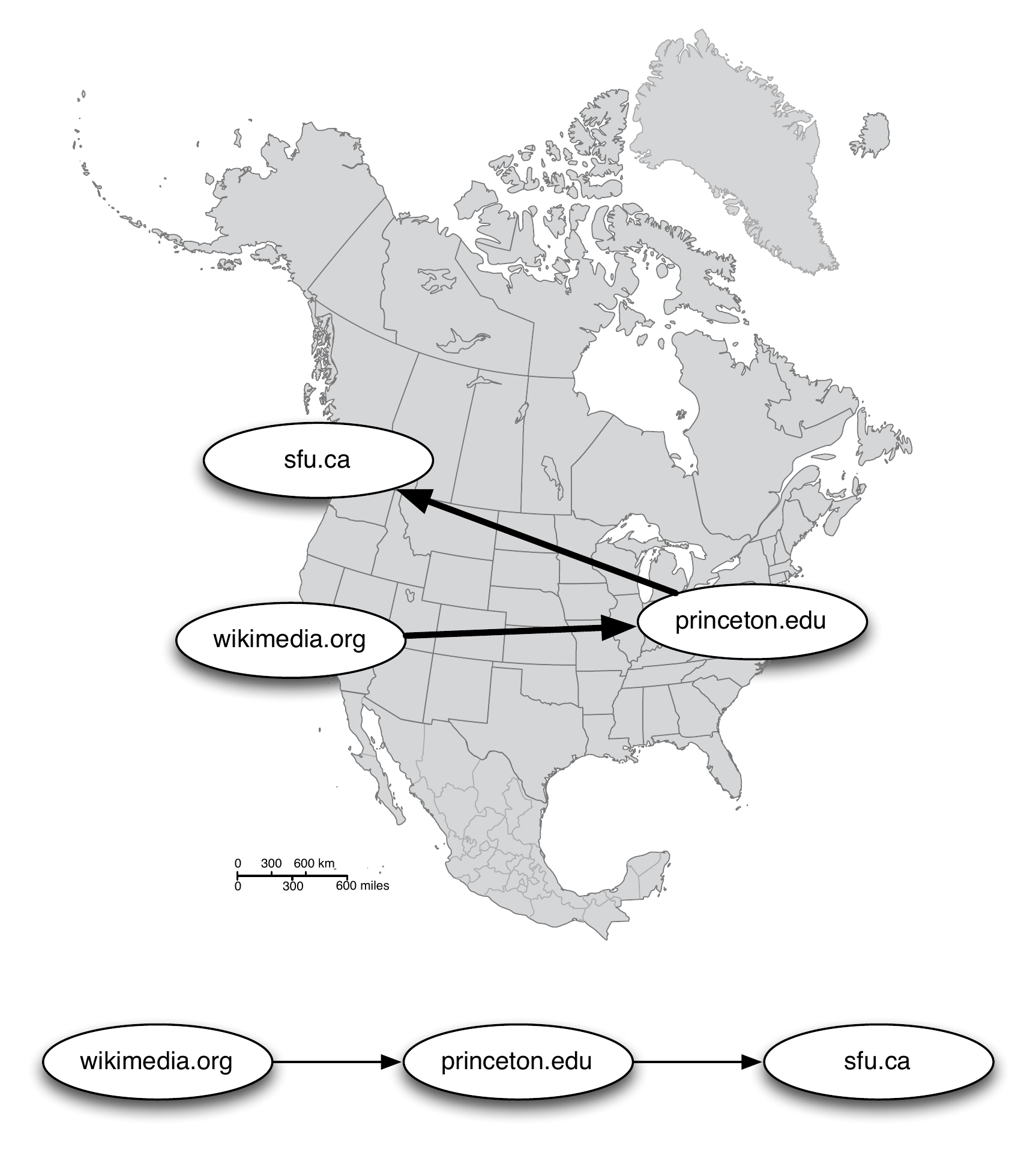}
\caption{Example workflow}
\label{fig:example_workflow}
\end{figure}

Figures \ref{fig:example_workflow} show a simple, sequential workflow with a data source (wikimedia.org) and two workflow nodes hosted on PlanetLab (\texttt{princeton.edu}, \texttt{surrey.sfu.ca}). It takes an image from \texttt{wikimedia} and then sends it to the \texttt{princeton.edu} node for processing. The result from this step is then sent to the \texttt{sfu.ca} node.  
\begin{figure}[!htp]
\centering
\includegraphics[width=2.2in]{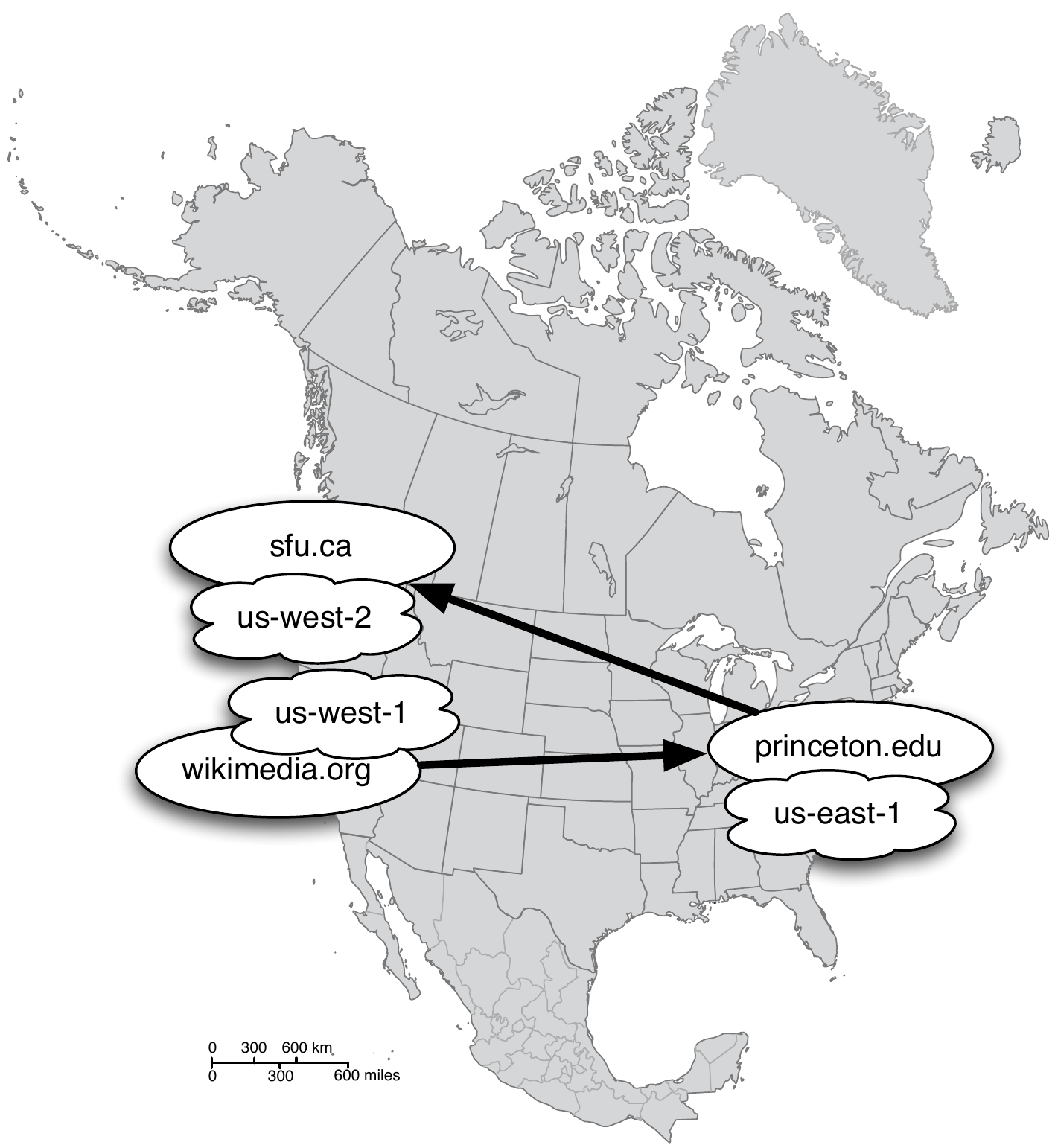}
\caption{Example workflow with closest Amazon EC2 regions}
\label{fig:example_workflow_with_ec2}
\end{figure}

\begin{figure}[!t]
\centering
\includegraphics[width=0.4\textwidth]{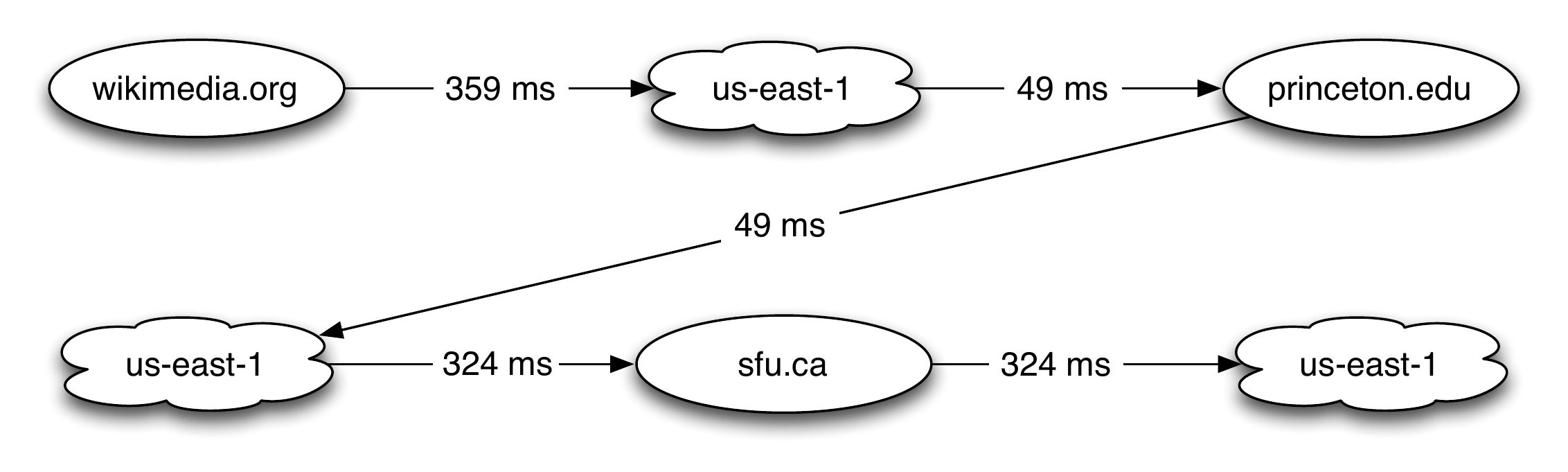}

\caption{Candidate workflow graph using EC2 region us-east-1 with network latency metric}
\label{fig:labeled_candidate_graph}
\end{figure}

Figure \ref{fig:example_workflow_with_ec2} shows the workflow illustrated in Figure \ref{fig:example_workflow} in relation to the closest Amazon EC2 Cloud regions (us-east-1, us-west-1, us-west-2). This illustrates the dilemma faced when deciding on the correct Cloud region to deploy the workflow orchestrator to: which Cloud region will result in the lowest execution time? 

CloudForecast attempts to solve this problem by evaluating the workflow based on: total geographical distance, network latency (as measured by the UNIX ``ping'' command) and HTTP round-trip time (as measured by the UNIX ``curl'' command) between every pair of adjacent nodes. It takes a workflow specification and builds several candidate workflow graphs, which represent the data flow. An example of a candidate graph for the workflow in Figure \ref{fig:example_workflow_with_ec2} can be seen in Figure \ref{fig:labeled_candidate_graph}. The tool further requires a list of Cloud regions (e.g. Amazon EC2 regions) that should be considered for workflow execution. Every candidate graph will be based on a different one of these Cloud regions and represent a possible optimised workflow using the Cloud instance as the workflow orchestrator. Every edge represents the data flowing from a Web service node to the Cloud region or vice versa. Since there are $8$ EC2 Cloud regions and $3$ metrics, the analysis tool will generate a total of $24$ candidate graphs.

Based on every candidate graph and metric used, CloudForecast computes an overall score. Using the geographical distance, the $n$ shortest Cloud regions are selected. Then, these are evaluated using the ``ping'' and HTTP RTT metrics and a final score is calculated. The final output is a table which ranks the Cloud regions by their predicted execution times. Table \ref{fig:example_result_table} is an example of the output table for the workflow displayed in Figure \ref{fig:example_workflow_with_ec2}.


 \begin{table}[!h]
 \renewcommand{\arraystretch}{1.2}
 \caption{Analysis tool result output}
 \label{fig:example_result_table}
 \centering
 \begin{tabular}{|l|l|}
 \hline
 EC2 endpoint & final score\\
 \hline
 us-east-1 & 92530.42\\
 us-west-2 & 186251.487\\
 us-west-1 & 186374.351\\
 sa-east-1 & 366450.152\\
 ap-northeast-1 & 421102.237\\
 ap-northeast-2 & 510982.726 \\
 ap-southeast-1 & 532180.129\\
 eu-west-1 &500178094.532\\
 \hline
 \end{tabular}
 \end{table}

\section{Experiments}
To verify the functionality of CloudForecast, we randomly generated workflows, analysed them and then executed and timed them several times. We chose to use randomly generated workflows to simulate real-life applications of workflows. Both simple sequential (A -- G) as well as more complex workflows (H -- I) involving sequential, fan-in and fan-out patterns with 2, 3, 4, 5, 7, 10, 12 and 13 nodes were generated.

In order to verify the optimisations proposed by CloudForecast, we implemented a very simplified, bare-bones workflow orchestrator which uses a set of custom workflow nodes implemented on top of PlanetLab.

\begin{figure}[h]
\centering
\includegraphics[width=0.4\textwidth]{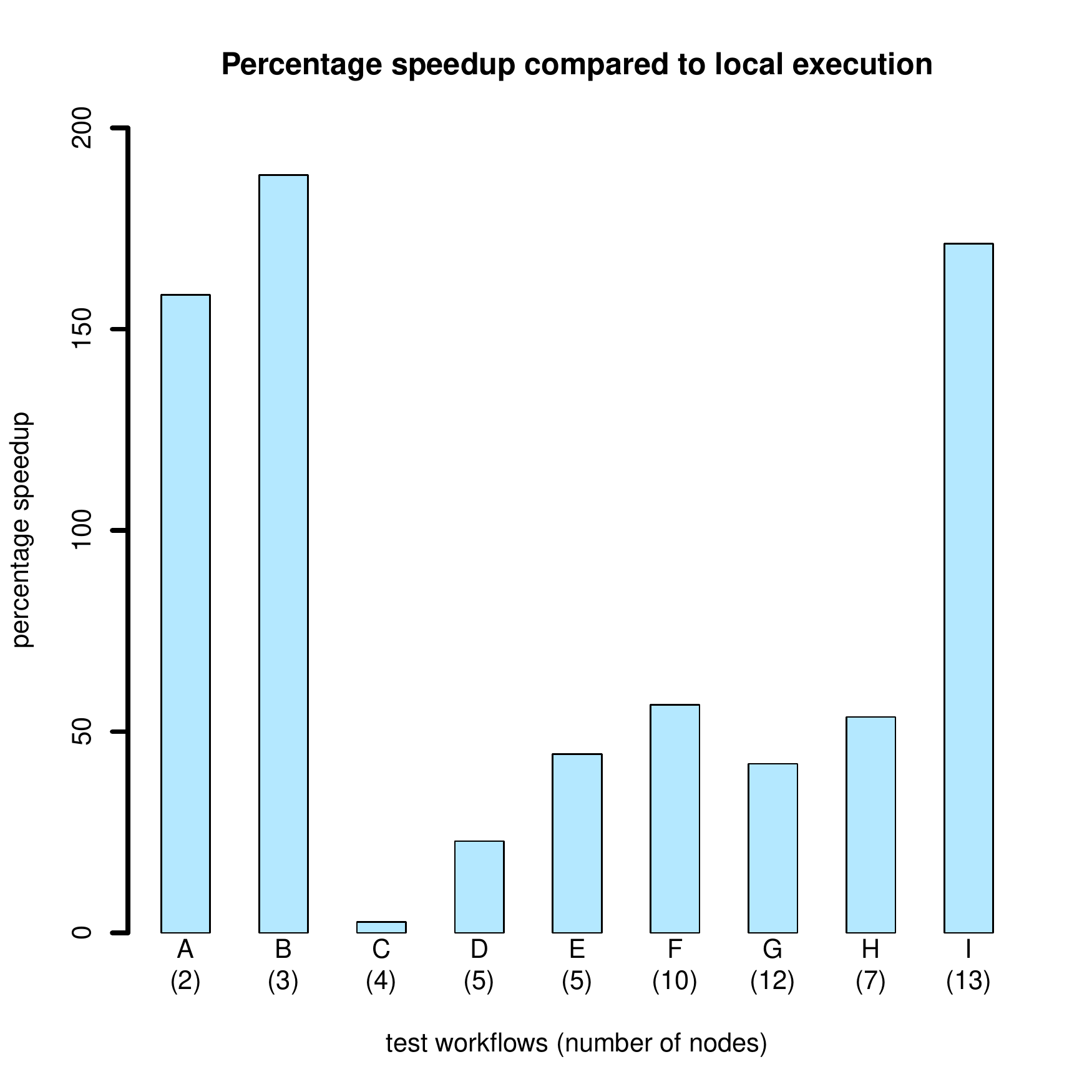}

\caption{Overall performance gain due to pre-deployment analysis}
\label{fig:overall_performance_gain}
\end{figure}

Figure \ref{fig:overall_performance_gain} illustrates the speedup in mean execution time for each sample workflow due to being run in the first-ranked Cloud region compared to local execution. The speedups range from $3\%$ to $188\%$ with a mean of $82.25\%$. We can conclude that the analysis correctly ranks the Cloud regions to reduce execution time. Please refer to \cite{DBLP:conf/cloudcom/LuckenederB13} for a more thorough presentation of the framework and evaluation. 

\section{Conclusions}
We discussed how to increase the performance of highly distributed Web service workflows by dynamically deploying the workflow orchestrator on an IaaS Cloud rather than orchestrating remote services locally. We conclude that orchestrating workflows in the Cloud significantly reduced overall execution time.

 \bibliographystyle{abbrv}
 \bibliography{cloud} 

\end{document}